\documentclass[onecolumn,authoryear]{./lib/els-mrw} 

\usepackage{amsmath,amssymb,amsfonts,amsthm,makeidx,graphicx}
\usepackage{txfonts}
\usepackage{helvet}
\usepackage{cleveref}

\newcommand{\MJup}{\mbox{${M}_{\rm Jup}$}}
\newcommand{\MMoon}{\mbox{${M}_{\rm Moon}$}}
\newcommand{\MSun}{\mbox{${M}_\odot$}}

\newcommand{\RSun}{\mbox{${R}_\odot$}}

\newcommand{\jumbo}{\mbox{JuMBO}}
\newcommand{\jumbos}{\mbox{JuMBOs}}

\newcommand{\Jumbos}{\mbox{JuMBOs}}

\def\apgt{\ {\raise-.5ex\hbox{$\buildrel>\over\sim$}}\ }
\def\aplt{\ {\raise-.5ex\hbox{$\buildrel<\over\sim$}}\ }
\def\lteq{\ {\raise-.5ex\hbox{$\buildrel<\over-$}}\ }
\def\lsquo{\ {'}\ }

\def\ni{\noindent}
\def\nl{\par\noindent}

\def\ind#1#2{\nl\hbox{\hskip 0.0truein\hbox to 0.3truein{#1\hfil}
	        \hbox{\hsize 6.4truein\vtop{\ni#2}\hfil}}\vskip -0pt}

\def\aap{\ {A\&A}\ }

\def\aj{\ {AJ}\ }

\def\apj{\ {ApJ}\ }
\def\apjl{\ {ApJL}\ }
\def\apjs{\ {ApJS}\ }
\def\apss{\ {Ap\&SS}\ }
\def\araa{\ {ARA\&A}\ }

\def\mnras{\ {MNRAS}\ }

\def\nat{\ {Nat}\ }

\def\psj{\ {The Planetary Science J.}\ }

\def\sovast{\ {SvA}\ }

\begin{document}

\chapter{The origin of free-floating objects in the Galaxy}\label{chap1}

\author[1]{Simon Portegies Zwart}%

\address[1]{\orgname{Leiden University}, \orgdiv{Leiden Observatory}, \orgaddress{Einsteinweg 55,
NL-2333 CC Leiden,
The Netherlands}}

\articletag{First version}

\maketitle

\begin{glossary}[Glossary]
\term{Cluster} A bound conglomeration of stars, planets, and other
minor bodies. \\
\term{JuMBO} Jupiter-Mass Binary Object.\\
\term{Naked star} is a single star, not orbited by a disk, planets or other minor bodies.\\
\term{Planet} A self-gravitating body with a mass $\apgt 0.01$\,\MMoon, but less massive than 50\,\MJup. \\
\term{Planetesimal bullying} the process in which minor bodies (planetesimal) are violently accelerated through gravitational assist of a much more massive body (star or planet). \\
\term{Relaxation:} The virialization of a bound stellar cluster, resulting in the velocity distribution approaching a Maxwellian. \\
\term{Singleton} or {\bf Dressed star}
is a single star, possibly orbited by planets and other minor bodies.\\
\term{Solus Lapis} or {\bf free-floating planetesimal} is a
minor body (with a mass between $10^3$\,kg and $\sim 0.01\,$M$_{\rm moon})$ not bound to a star or planet. A Solus lapis could be bound to a cluster of stars.\\
\term{Star} A Hydrogen or Deuterium burning body with
a typical mass $\apgt 0.05$\,\MSun.\\
\term{Virial:} The equilibrium energy state of the stellar cluster,
referring to the virial theorem.  In stellar dynamics often referring
to the ratio of the kinetic to the potential energy to be $0.5$. \\

\end{glossary}


\begin{abstract}
  The Milky way Galaxy is brimming with free-floating objects,
  including stars, planets and planetesimals.  For the purpose of this
  chapter, we define a free-floating object as a solid body that is
  not orbited by a considerably more massive body. A planet then is
  considered free floating if it is not orbiting a star but it may
  be orbiting another planet. A binary planet, or planet-moon pair
  that is not orbiting a star, is then considered free floating.

  Most free-floating objects are not born as such because most
  objects form in some sort of coordinated environmental effort,
  such as a star forming region or a circum-stellar disk.

  free-floating stars then originate from dissolved clusters.  Free
  floating planets are ejected from their parent star in an internal
  dynamical encounter with another planet or stripped from the star by
  other means such as a supernova or a nearby passing star. Free
  floating (interstellar) planetesimals probably form in a similar
  fashion as free-floating planets.

  The number of free-floating objects in the Galaxy can be large.
  With billions of stars and planets, and trillions of interstellar
  planetesimals. Although free-floating planets appear to be quite
  common (a few hundred have been observed), only two interstellar
  planetesimals have been discovered so far. The expectation, however,
  is that they outnumber the stars in the Galaxy by a considerable
  margin. We expect them to be found more frequently once large new
  instruments come online, such as the Vera Cooper Rubin Observatory.
\end{abstract}

\section*{Key elements}
\begin{itemize}
\item[$\bullet$] Free-floating planet-mass objects ($m \apgt
  0.01$\,M$_{\rm moon}$) in the Milky way Galaxy outnumber stars by up to a
  factor of $3$.
\item[$\bullet$] Young star clusters ($\aplt 1$\,Myr) host
  approximately one free-floating planet pair per $4$ stars, whereas
  they are absent in older clusters.
\item[$\bullet$] Free-floating planetesimal-mass ($m \aplt
      0.01$\,M$_{\rm moon}$) objects in the Milky way Galaxy outnumber stars
      by up to a factor of $10^{15}$.
\item[$\bullet$] The mass distribution of free-floating minor bodies
  appears to be flatter than the stellar-mass distribution.
\end{itemize}
  
\section{Introduction}\label{chap1:sec1}

The majority of the Milky Way Galaxy's mass is composed of dark
matter, stars, and interstellar material (gas and dust), likely in
that order of importance in terms of their mass contribution; and then
there is a lot of stuff we don't see, such as pebbles.  Most stars in
the Galaxy have companions.  There may be a very small population of
stars without any companions (naked stars), and if so, there is
generally a good evolutionary reason why they are not dressed.
Accompanied stars are orbited by other stars (forming binaries or
multiple systems), or lower-mass objects such as planets, minor
planets, planetesimals, and comets.  Planets themselves can be orbited
again by moons or rings.  So far there seems no bottom to this
hierarchy, as planetesimals appear to have moons, or even comparable
mass companions. Binary planetesimals appear particularly abundant in
the Kuiper belt, where the binarity exceeds 20 per cent
\citep{2024PSJ.....5..143P}\footnote{The author had to limit the
number of references to 45, which was close to impossible. However, he
managed to bring it down from 150 to 68, which was accepted by the
editor. Quite some important references are therefore missing, but
only the most onvious ones are left out.}.  Even rings could be
common, as a few have already been observed, such as the ring around
in the Centaur (10199) Chariklo.
A similar argument could hold for moons, but these have so far not
been found orbited by (sub)moons, planetesimals or rings
\citep{2019MNRAS.483L..80K}\footnote{The planets Kepler-1625b and
Kepler-1708b, however, could, from a theoretical perspective, be
orbited by moons that are orbited by moons
\citep{2023MNRAS.520.2163M}. Therefore, free-floating sub-moons could
exist.}.

The interactions among these bodies lead to the dynamic evolution of
the planetary systems.  The reorganization of minor bodies is
primarily driven by giant planets. During their dynamic interactions,
the orbits of minor bodies are reshaped, and objects change
role. Moons may become planets if dislodged from their parent planet,
and planetesimals can collide with each other and with other
bodies. New moons or rings may form through giant impacts, and objects
may be ejected from the planetary system. Apart from the complexities
in dynamical evolution, the challenge of keeping track of the
evolution of a planetary systems is further complicated by the
evolving nomenclature.

The picture of a Galaxy with stars orbited by planets, and planets
orbited by moons becomes more intricate with the realization that
these systems interact both internally and with each other. Such
interactions expand the range of possible outcomes even further. In
clusters, planetary systems may interact and exchange material, or
material may escape the gravitational pull of its parent star, leading
to free-floating objects.  

Among the free-floating minor bodies in the Galaxy, planetesimals and
comets are probably the most common, followed by moons and
planets. Observationally, distinguishing free-floating planets from
moons can be challenging. To simplify nomenclature, we based the
definition on the current objects' whereabouts and disposition.  We
will often reference the Solar System as a typical planetary
system. Although it is probably not representative, it remains the
closest and best-studied example.

This chapter discusses free-floating objects in the Milky Way Galaxy,
following a mass hierarchy: stars, and planets, then planetesimals and
comets. We begin with an observational perspective and proceed with
more speculative theoretical arguments.

\section{Observational perspective}\label{chap1:Observations}

\subsection{free-floating stars}

\begin{definition}
A free-floating star is a star not directly bound to any other star or
member of a bound group of stars.\footnote{Note here that these
definitions are not necessarily consistent with the IAU's definition,
and that they reflect the author's opinion.}
\end{definition}

Stars are born clustered \citep{2010ARA&A..48..431P}.  The majority of
these clusters dissolve again leaving their constituents as isolated
objects in the Galaxy.  Roughly half to 90 per cent of the stars are
born in non-virial or fractal-structured environments
\citep{2022MNRAS.515.2266A} that experience violent relaxation in the
first 10\,Myr.
Such a cluster
dissolves shortly thereafter.
The total fraction of stars in bound open clusters today is about $1$
to $5$ per cent \citep{2024A&A...686A..42H}.  Globular clusters
constitute about 1 per cent of the stars in the Galaxy.  The vast
majority of stars then are singletons, or the member of a binary or a
multiple system.  Most of these, however, are probably dressed,
i.e. orbited by planets and other minor bodies.  The Sun is such a
star; it is free-floating, dressed with planets, minor planets,
planetesimals, and comets.  Undressed free-floating stars are probably
rare, but we can only speculate on the absence of planets around
singletons, as it is hard to prove a star to be truly single. We dwell
a bit on the theoretical arguments for naked free-floating stars in
\cref{chap1:ffstars}.

If stars are considered free-floating when they are isolated, their
evolutionary products can be as well.  Most neutron stars are probably
singletons.
The enormous mass loss and high-velocity kick during their transition
from star to neutron-star almost guarantees them to be single. Black
holes, equally wise, are quite likely to be free floating; although
they are unlikely to receive a high-velocity natal kick in the
supernova.
An example of a free-floating black hole, discovered through
microlensing, is MOA-11-191/OGLE-11-462
\citep{2022ApJ...933...83S}. This black hole, with a mass of
$7.1\pm1.3$\,\MSun\, and located at a distance of $1.58\pm0.18$\,kpc,
has no detectable optical counterpart.

\subsection{free-floating planet-mass objects}\label{chap1:Obs_ffplanets}

\begin{definition}
A free-floating planet is a planet not bound to a star or orbiting a
multiple system.
\end{definition}

The first direct imaged free-floating Jupiter-mass objects were
discovered in the direction of the Trapezium cluster more than twenty
years ago
\citep{2000Sci...290..103Z}.
Since then, many more have been found, for example, in the young
clustered environment of Upper Scorpius \citep{2022NatAs...6...89M},
and through gravitational microlensing surveys in the direction of the
Galactic bulge \citep{2012ARA&A..50..411G}.  Their
abundance may be between $0.25$ per star \citep{2017Natur.548..183M}
and $1.9^{+1.3}_{-0.8}$ per star \citep{2011Natur.473..349S}, although
a considerable fraction of these could be in wide ($\gg 100$\,au)
orbits around a parent star, or have masses $\ll \MJup$.

\begin{figure}[t]
  \centering
  \includegraphics[width=1.0\linewidth,trim={1cm 0cm 0cm 1.0cm}]{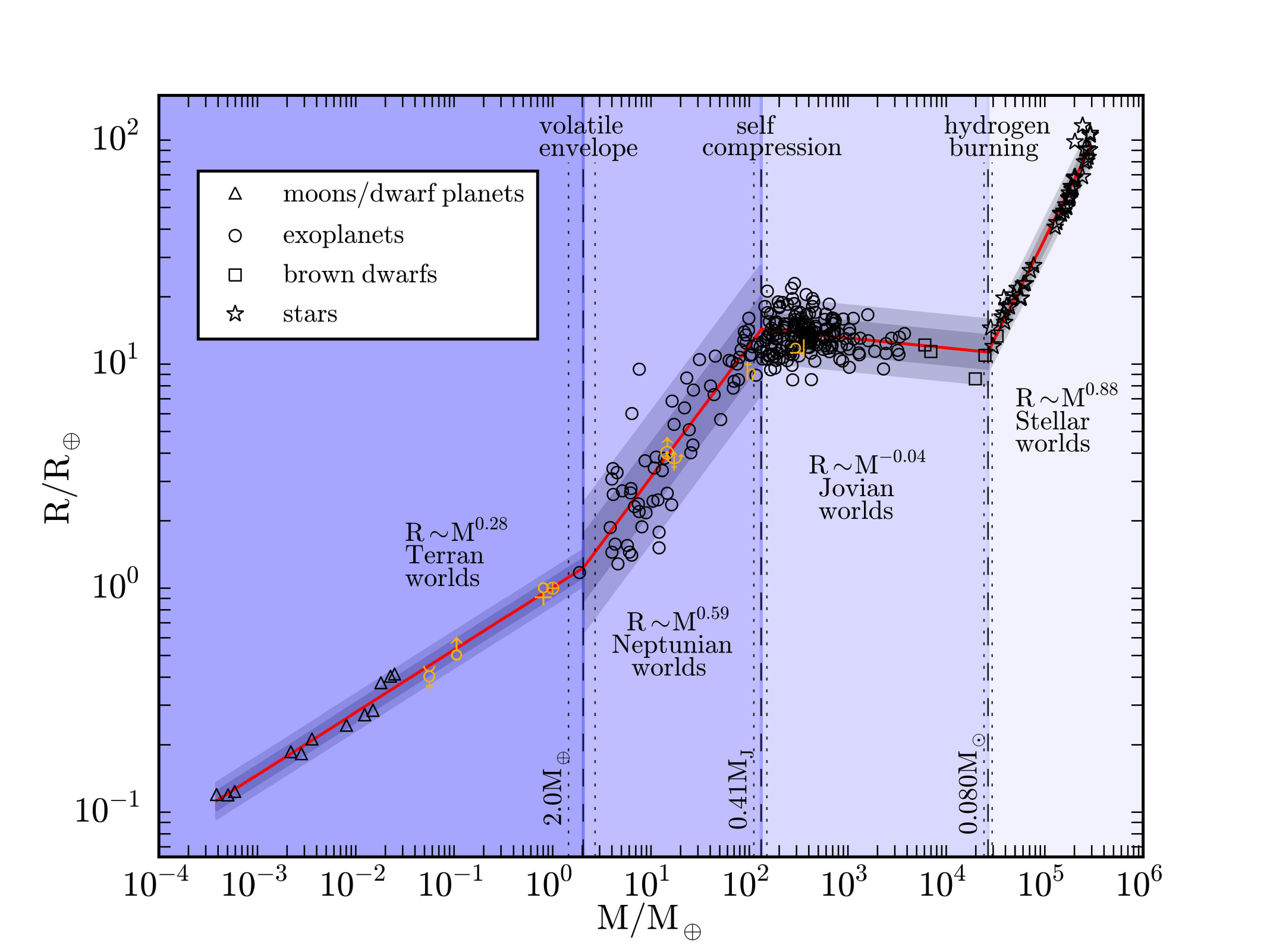}
\caption{Mass-radius relation for planets and low-mass stars, from
  \cite{2017ApJ...834...17C}.  The symbols (identified in the top
  left) represent observed systems, with a probabilistic model
  overplotted with lines \citep[and 68\% and 95\% conﬁdence intervals
    in gray,][]{2017ApJ...834...17C}.  }
\label{fig:planet_or_star}
\end{figure}

Discovering free-floating planets is complicated by their lack of an
energy source on their own. At young age, low-mass stars and high-mass
planets are hardly distinct, making it difficult to identify a planet
from a star, see however \cite{2005AGUFM.P11A0100D} for a tentative
definition of a planet.  The problem in recognizing planets from stars
is probably best illustrated in fig.3 of \citep{2017ApJ...834...17C};
for convenience reproduced as \cref{fig:planet_or_star}.  Here they
show the complex mass-radius relation for observed exoplanets, with
various ``families'' of planets identified.

The mass-radius relation and the mass-temperature relation differ for
planets and for low-mass stars.  Such a relation can be derived by
combining 1-dimensional radiative-convective equilibrium atmospheric
models, with the equation of state for dense hydrogen-helium mixtures
\citep[see][]{2017ApJ...834...17C}.  Further complications are
introduced by the adopted (non)equilibrium chemistry combined with
parameterized relations for eddy diffusion with surface gravity for
planet atmospheres.

To circum navigate this star-planet delimitation discussion, we
mention the first three discovered free-floating planets, while
keeping in mind that they may actually be brown dwarfs.  The first
direct imaged free-floating planet was S~Ori~52 in the $\sim 3$\,Myr
old cluster $\sigma$~Orionis at a distance of $391^{+50}_{-40}$\, pc
\citep{2022yCat.1355....0G}\footnote{Interesting to note is that this
object is considered a free-floating planet, even though it is
probably bound to the cluster.}.  With a mass of $5$ to $15$\,$M_{\rm
  Jupiter}$ \citep{2000Sci...290..103Z}, S~Ori~52 is probably a true
planet, i.e. there is no Deuterium fusion\footnote{Deuterium fusion
becomes possible for masses around 13 to 60\,\MJup\,
\citep{2016Ap&SS.361...89O}.} (but see below).  The young age of the
cluster makes its population of free-floating planets visible in the
infrared. Such planets still contract from their formation process and
therefore have a surface temperature of some 2000\,K, sufficiently
warm to be observed with infrared telescopes.

\begin{figure}[t]
\centering
\includegraphics[width=1.0\textwidth]{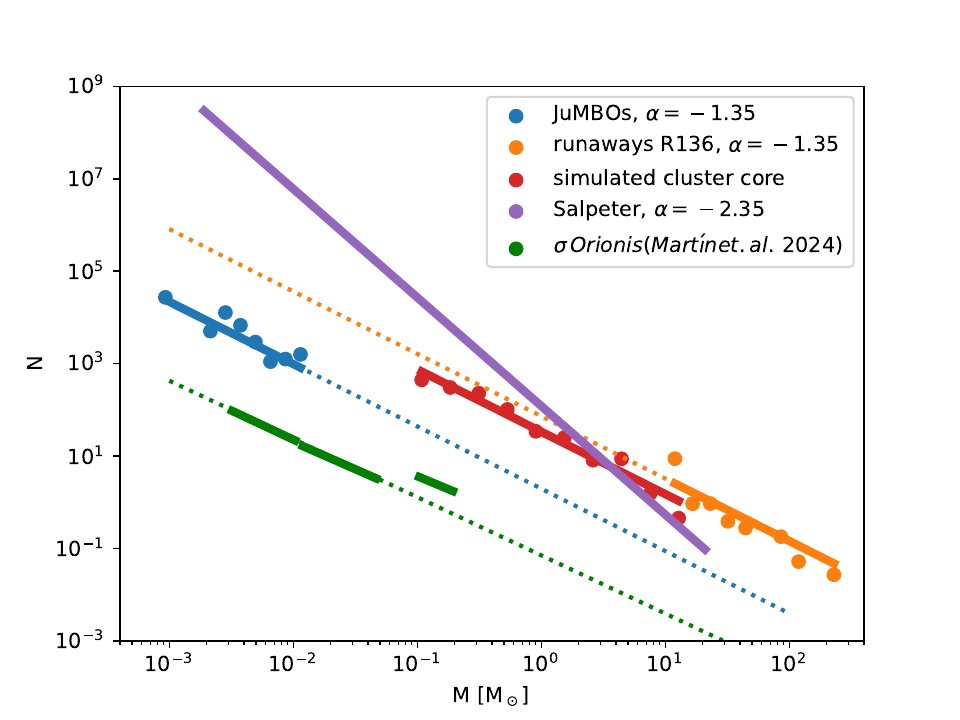}
\caption{The mass function for stars and planet-mass objects.  The
  Salpeter mass function (purple line) is presented to highlight the
  difference with the other mass functions. We also present the mass
  function of the observed runaway stars in the Tarantula nebula
  \cite[][in orange]{MStoopetal2024}, the mass function of the
  observed primaries of \jumbos\, in the Trapezium cluster
  \cite[][blue]{2023arXiv231001231P}, and the planet-mass objects
  found in the $\sigma$\, Orionis cluster \citep{2024arXiv240513497M}
  (greed).
  For comparison, we also present the mass function for the core
  population of a simulated cluster of 6890 stars \cite[data
    from][]{WilhelmandPZ2024}. For the latter we continued the evolution
  of the cluster until the moment of core collapse, of which the
  core-mass function is presented here. To guide the eye, we present
  the dotted curves, which are analytic extension of the measured mass
  functions with the slope indicated in the legend.  }
\label{fig:ff_massfunction}
\end{figure}

Recent observations with the Euclid telescope confirms such detections
by finding more than 90 free-floating planets down to $4$\,\MJup\, in
$\sigma$\,Orionis \citep{2024arXiv240513497M}.  They also see
S~Ori~52, but identify it as an L0.5 spectral-class brown dwarf.  In
total they identify some 95 objects with a mass below $0.1$\,\MSun\,
with an average power-law mass function of $\alpha \simeq -1.2$.
Interestingly enough, \cite{2021ApJS..253....7K} found the sub-stellar
mass function to have a somewhat steeper average power-law slope of
-1.6.  In \cref{fig:ff_massfunction} we present their measured mass
function as the green line (and the Salpter mass function in
purple)\footnote{We adopt the mass function of the form $dN/dM \propto
M^\alpha$, in which case for Salpeter $\alpha = -2.35$.}

The first Spitzer-detected free-floating planet is the $\sim
11.5$\,\MJup\, object OTS 44 \citep{2005ApJ...620L..51L}.  The first
microlensing free-floating planets are OGLE-2012-BLG-1323 ($0.0072$ to
$0.072$\,\MJup) and OGLE-2017-BLG-0560 \citep[$1.9$ to
  $2.0$\,\MJup,][]{2019A&A...622A.201M}.  The mass estimate of the
objects detected through microlensing is generally degenerate
\citep[][we only list the lower bound]{2012ARA&A..50..411G}. These
objects are often distant and sighted only once; it is unlikely that
they will be seen again (at least not for another Galactic orbit).

Recently \cite{2023arXiv231001231P} reported the discovery of 540
single Jupiter-mass objects with a mass between 0.6\,\MJup\, and
14\,\MJup\, in the direction of the Trapezium cluster.  Among these
are 42 projected pairs (called JuMBOs) with projected separations
between 25\,au and 380\,au, and two with a Jupiter-mass tertiary
companion.

The mass function of \jumbo\, primaries is rather flat.  In
\cref{fig:ff_massfunction} we present the mass function for runaway
stars, and the primary masses of the \jumbos\, observed in the
direction of the Trapezium cluster. Interestingly \jumbo\, masses have
a similar slope as the runaway stars and free-floating planets
observed elsewhere. The latter is a dynamical population, formed
through the dynamical sling-shots in multi-body interactions in the
mass-segregated core of a stellar cluster. \Jumbos, on the other hand,
seem to have formed in situ (see \cref{chap1:ffplanets}), and there
are compelling reasons to expect their mass function to be flat,
because the brown-dwarf mass-function, down to $\sim 5$\,\MJup\, is
flat \citep[$\alpha \simeq 1.5$,][]{2024arXiv240609690R}.

Apart from the observed JuMBOs in the direction of the Trapezium cluster, only four other paired planetary-mass objects are known:
\begin{itemize}
\item[$\bullet$]2MASS J11193254-1137466 AB: a $5$ to 10\,\MJup\,
  primary in a $a=3.6\pm0.9$\,au orbit \citep{2017ApJ...843L...4B}.
\item[$\bullet$]WISE 1828+2650: a 3 to 6\MJup\, primary with a
  5\,\MJup\ companion in an $\apgt 0.5$\,au orbit
  \citep{2013ApJ...764..101B}.
\item[$\bullet$] WISE J0336-014: a $8.5$ to
  $18$\,\MJup\ primary with a $5$ to $11.5$\,\MJup\, companion in a
  $0.9^{+0.05}_{-0.09}$\,au orbit \citep{2023ApJ...947L..30C}.
\item[$\bullet$]2MASS J0013-1143
  is suspected to be a binary by \citep{2019A&A...629A.145E}.
\end{itemize}
Each of these was found with the IRSA Two Micron All Sky Survey
(2MASS) or by the space-based Wide-field Infrared Survey Explorer.

With the launch of the JWST and Euclid, the population of directly
imaged free-floating planet-mass objects is already increasing
dramatically.  With the recent discovery of 540
free-floating planet-mass objects by JWST in the Trapezium cluster
\citep{2023arXiv231001231P}, and more than $10^5$ low-mass (unresolved
red) objects in the star-forming region M78 by Euclid
\citep{2024arXiv240513497M}, this trend has already set in.

\subsection{Interstellar free-floating objects}\label{chap1:S.SolusLapis}

\begin{definition}
  A free-floating planetesimal is a rocky or icy object not
  gravitationally bound or orbiting a planet or star.
\end{definition}

The population of Solus Lapides
\citep{2018MNRAS.479L..17P}\footnote{Latin for {\rm lonely stone}.}
has gained considerable traction since the discovery of the
interstellar interlopers: 'Oumuamua \citep{2017Natur.552..378M}
and Borisov \citep{2021SoSyR..55..124B}.  Both object entered the
Solar system on an hyperbolic (unbound) orbit with a closest approach
to the Sun of 0.256\,au and 2.0\,au, for 'Oumuamua and Borisov,
respectively.  As a consequence, they can be observed only when
passing through the Solar system, which happens only once.  By the
absence of any other objects in this classification the discussion is
somewhat theoretical.  Both objects are currently beyond telescope
range, with a visual magnitude for 'Oumuamua of $m_v \apgt 38.5$.

\paragraph{1I/'Oumuamua}
This object was discovered on October 19, 2017, by the
Pan-STARRS1 telescope in Hawaii \citep{2017Natur.552..378M}. The "1I"
in its official designation indicates that it is the first known
interstellar object.  The term 'Oumuamua, refers to "messenger" in
Hawaiian; In my opinion it could have also been aptly named
Rama\footnote{After the 1973 book ``Rendezvous with Rama'' by
A.C.\,Clarke.}, but politics and religion were in the way.  It came
from the direction of the star Vega in the constellation Lyra and is
now heading roughly in the direction of $\psi$ Pegasus.  'Oumuamua is
currently at a distance of 6.3 billion km ($\sim 42$\,au) from the Sun
moving away at a speed of 17km/s (or $\sim 5.4 \times
10^{27}$\,\r{A}/Myr).

'Oumuamua is peculiar in many aspects. It is highly elongated, with
estimates suggesting it could be 800 meters long and about 80 meters
wide. It appears to be tumbling with an 8-hour period, rather than
spinning on a single axis \citep{2017Natur.552..378M}.
Its reddish color suggests the presence of organic-rich material on
its surface.
Unlike typical comets, 'Oumuamua showed no sign of a coma.

The origin and nature of 'Oumuamua have been actively debated. It may
have come from a nearby stellar association \citep{2018ApJ...852L..27F}
or roamed the Galaxy for the last several billion years
\citep{2018MNRAS.479L..17P}. This could be explained by its interaction
with interstellar sputtering (blackening it's surface)
\citep{2017RNAAS...1...50D} or being a fragment of a larger rocky
object \citep{2018arXiv180202273K}, a shard ejected from a stellar
binary \citep{2018ApJ...852L..15C}, or a tidally destroyed comet
\citep{2020NatAs...4..852Z}.

The orbit of 'Oumuamua has been of particular interest as it seemed to
accelerate away from the Sun after perihelion passage
\citep{2018Natur.559..223M}.
With the lack of detectable CO or CO$_2$ this acceleration cannot be
attributed to a comet-like outgassing.  There are quite a number of
alternative explanations, though, including N$_2$ or H$_2$O
outgassing, which would be hard to observe,
radiation-driven acceleration,
nozzle-like venting of volatiles,
or other more exotic mechanisms.  If one favors the hypothesis that
'Oumuamua is an crafted object by an extra-solar civilization, I would
favor its similarity with the Troja space craft\footnote{Referring to
the 1982 Perry Rhodan \#233 ``Geheimsatellit Troja'' by
K.H. Scheer}. In that case, its occupants ware not interested in the
Solar system's inhabitants because of our low Kardashev scale
\citep[type I,][]{1964SvA.....8..217K}.

It seems most plausible that 'Oumuamua is a leftover planetesimal from
a planet-forming disk, ejected by a giant planet through gravitational
assist \citep{2019NatAs...3..594O}, or a former Oort cloud object
\citep{2019AJ....157...86M}. In that case, an Oort-cloud formation
efficiency of a few per cent, what seems reasonable for the Solar
System, could have produced around $10^{12}$ such objects during the
formation of the giant planets and the ejection of planetesimals into
the Oort cloud.  Before passing through the Solar system, it probably
roamed the Galaxy for several billion years
\citep{2019NatAs...3..594O}.

With a galactic mean density of $10^{14}$ or $10^{15}$ of such objects
per cubic parsec \citep[as derived by][based on a single
  observation]{2018MNRAS.479L..17P}, the rate of capture by the Solar
System is estimated at $2 \times 10^{3}$ per million years
\citep{2022MNRAS.512.4062D}. The rate at which they are accreted by
the Sun is about an order of magnitude higher
\citep{2022MNRAS.512.4078D}. The total amount of interstellar
asteroids accreted over the Sun's lifetime then amount to roughly the
mass of the Moon, which, asteroids being rich in Platinum group
elements, could have enriched the Sun by about one part-per-million.

\paragraph{2I/Borisov}
This object was discovered by amateur astronomer Gennadiy Borisov on
August 30, 2019 \citep{2021SoSyR..55..124B}.  Borisov has a visible
coma, and a tail composed of dust and gas,
which is unusually rich in CO, CN \citep{2020NatAs...4..861C},
and potentially even some water
\citep{2020ApJ...889L..10M}.  The reporting of outbursts by
suggests that also the interior may be ice-rich.  The spherical
equivalent radius of the nucleus is at most half a kilometer
\citep{2020ApJ...888L..23J}, making it potentially comparable in size
as 'Oumuamua.  Borisov came from the direction of the constellation
Cassiopeia, passing the M0V star Ross 573 some $\sim 0.91$\,Myr ago
within a distance of 0.068\,pc \citep[$\sim
  14000$\,au,][]{2020A&A...634A..14B}.  Borisov is currently traveling
through the Solar system on a hyperbolic trajectory in the direction
of the star $\beta$~Ara.

Unlike 'Oumuamua, Borisov has been observed to behave very similarly
to typical solar-system comets, providing valuable data on the
composition of interstellar objects.  Although
\citep{2018MNRAS.479L..17P} predicted, upon the discovery of
'Oumuamua, such objects would be found regularly (on a yearly basis
given current telescope sky-coverage), their euphoria upon the
discovery of Borisov was soon thereafter quenched by the lack of
further new discoveries since 2019.

\section{Formed bound to become free floating}\label{Sect:Environment}

To appreciate the theoretical discussion on free-floating objects in
\cref{chap1:ffstars}, we start by discussing the formation and early
evolution of bound conglomerates, such as stellar clusters and
planetary systems.

The Sun is dressed with planets and planetesimals, but it is not
accompanied by other stars. The Sun, then, hovers in the Galactic
potential as a free-floating star. For stars in the neighborhood this
seems to be the norm rather than the exception\footnote{The solar
neighborhood is ill defined, but here we adopt all stars within
${\cal O}(10)$\,pc. According to ChatGPT: {\em Typically, the Solar
    neighborhood is considered to include stars within about 20 to 50
    light-years from the Sun.}}.  To prevent the history and
observational bias to interfere with our discussion, we consider three
categories of free-floating objects: 1: stellar-mass objects, 2:
planet-mass objects, and 3: sub-planet mass objects. We do not
consider dust or pebbles, but they could be considered to be part of
the interstellar medium.

For clarity, we then have the definition based only on mass, shape,
and energy source.  Stars acquire their energy from Hydrogen or
Deuterium burning. Planets acquire their energy from contraction, and
are held together by gravity.  Planetesimals are held together by
either (or both) gravity or material strength, but do not generate
energy by contraction. Pebbles are the building blocks of
planetesimals. One can then roughly summarize in terms of mass as
follows: Stars are more massive than $\sim 0.05$\,\MSun\, ($\sim
50$\,\MJup), planets are more massive than $10^{20}$\,kg (around the
mass of the dwarf-planet 2~Pallas, or $0.01\,$M$_{\rm moon})$, and
planetesimals are more massive than $\sim 1000$\,kg. Smaller mass
objects are pebbles\footnote{A Harley Davidson FXBR motorcycle is a
pebble in this terminology.}.  As illustrated in
\cref{fig:planet_or_star}, we recognize a variety of sub-planet
classes, based on mass and composition. A similar discussion probably
holds for interstellar planetesimals, but so far we only have two
categories, dry (like 'Oumuamua) or wet (such as Borisov).

Before we delve into the various subcategories, it helps to sketch the
theoretical background, which includes the Milky Way Galaxy, the
stellar birth environment, and the planet forming disk around young
stars.  After that, in \cref{Sect:Theory} we continue the more
theoretically oriented discussion on the origin of the various
free-floating objects.

\subsection{Formation and evolution of the parent star cluster}\label{Sect:parentcluster}

Stars form through the self-gravitational collapse of a giant
molecular cloud.  The gaseous clumps become denser and hotter, and due
to the conservation of angular momentum each condensation acquires a
disk-shaped gaseous structure. This disk will later form the basis for
the planets (or possible stellar companions).

Although ill-understood, the star formation process is expected to
take place over a time frame of $\aplt 10$\,Myr.  Massive stars are
also less common, and their masses are distributed in a power-law with
a typical slope of -2.35 (the Salpeter mass-function, see
\cref{fig:ff_massfunction}).

One requires quite a lot of gas in a sufficiently small volume before
it becomes gravitationally unstable, and collapses into stars. This
amount of gas is called the Jeans mass, which approximately
\begin{eqnarray}
  M_{\rm J} &=& 4/3\pi \rho R_{\rm J}^3 \nonumber \\
  &\simeq& 2\,\MSun\, 
  \left({v_c^3 \over 0.2{\rm km/s}}\right) 
  \left({n\over 10^3 {\rm cm}^{-3}}\right)^{-1/2}.
\end{eqnarray}
Here $\rho$ is the stellar density in a volume with the Jeans radius
$R_{\rm J}$, $v_c$ is the sound speed, and $n$ is the number density,
with typical values already filled in. Stars then form in groupings,
or clusters.  This formation process is quite dynamic, but once the
molecular cloud runs out of gas, the remainder will be ejected or
blown away, leaving the left-over stellar distribution in a somewhat
super-virial state. The term ``somewhat'' here indicates our lack of
understanding this process. The gas is not homogeneously distributed,
with the central portion of the stellar distribution probably is gas
deprived.  As a consequence, the outer regions are more severely
effected by the outgassing compared to the inner bound central
region. Simulating this process is a daunting endeavor and actively
researched.

Once the gas is removed, the cluster evolution becomes dominated by
Newtonian dynamics and stellar mass loss.  Eventually, the cluster
dissolves. The evaporation of star clusters, yet another active field
of research, is driven by stellar mass loss in combination with
dynamical relaxation and the interaction with the Galactic tidal
field.
The evaporation process depends on the cluster density and the number
of stars, but generally lasts more than 100\,Myr; some of the most
massive star clusters are still around since the earliest times.

\subsection{The origin of free-floating stellar-mass objects}\label{chap1:ffstars}

\begin{definition}
A free-floating object is an object that is not orbiting any objects
more massive than itself.
\end{definition}

Stars are born in clusters through the process roughly summarized in
\cref{Sect:parentcluster}. The cluster disperses with time, and its
stars become singletons, with some possible paired or triple objects.
The time scale on which an isolated star cluster dissolves is related
to the relaxation time scale \citep[formally called the two-body
  relaxation time scale at the half-mass
  radius,][]{1987degc.book.....S}.
\begin{eqnarray}
	t_{\rm rlx} &= &\left( {R^3 \over G M} \right)^{1/2} {N \over 8
          \log\Lambda} \nonumber \\
        &\simeq&
        20\,{\rm Myr} \left({R \over 1 {\rm pc}}\right)^{3/2}
        \left( M \over {1000\MSun} \right)^{-1/2}
\end{eqnarray}
Here $G$, $R$, $N$, and $M$ are Newtons gravitational constant, the
cluster radius, the number of stars and their total mass,
respectively. The parameter $\log\Lambda \sim \log(0.4N)$ is the
coulomb logarithm.
Every relaxation time scale the velocity dispersion of the cluster
stars acquires a Maxwellian distribution, of which the tail exceeds
the escape speed. The time scale for an isolated cluster to dissolve
is then about $t_{\rm dis} \sim 15t_{\rm rlx}$
\citep{1987degc.book.....S}.  In the early 1950s Jan Oort did not
understand why the number of $\apgt 500$\,Myr old cluster was so small,
but it was only later that one realized that clusters are affected by
nearby giant molecular clouds and the Galactic tidal field.  In the
Galactic potential the cluster lifetime is shortened quite
dramatically, but the dependency on the mass, Galactic orbit and
cluster structure also becomes more complicated.
Both processes reduce the mean lifetime to a few 100\,Myr (with the
exception of globular clusters).

Once a star escapes a cluster, it becomes single\footnote{Note that
there is a finite probability that two (or more) stars capture each
other once the cluster potential diminishes, forming a pair (or
multiple) \citep{2011MNRAS.415.1179M}.}.  The way in which the star
escapes may have considerable repercussions for the orbiting minor
bodies.  The planets orbiting a star that is ejected through a
relatively strong encounter are more strongly affected than when the
star leaves the cluster because the cluster slowly dissolves in the
Galactic tidal field. In particular, the outer-most orbits can be
affected quite severely.  These perturbations can propagate to the
inner orbits, if the outer object is relatively massive or the
planetary system is fully packed.  Stars that are violently ejected
from their birth cluster become heavily perturbed and easily lose their
planets (and other minor bodies) in the process.

From a theoretical perspective free-floating stars are then a natural
consequence of the way star clusters evolve and disperse. Stars can
even lose their bound material when sufficiently accelerated to high
velocities, such as can occur when ejected through the gravitational
assist of a supermassive black hole,
or when the star experiences (or its companion) a supernova.
The first discovered hyper-velocity star, SDSS J090745.0+024507, was
probably ejected by the supermassive black hole on the Galactic center
with a velocity of $853\pm12$\,km/s \citep[][more than 28 times higher
  than the Earth's velocity in orbit around the
  Sun]{2005ApJ...622L..33B}.  A planet at a distance of 8\,\RSun\,
would barely have remained bound if this star was accelerated to the
observed velocity, and tigher orbits would end-up colliding with the
star.

Although so far, only one free-floating black hole was found
(MOA-11-191/OGLE-11-462), they must be quite common. Each $\apgt
10$\,\MSun\, high-velocity star (OB runaway star) eventually collapses
into a compact object in a supernova explosion, contributing to the
population of free-floating compact objects.

\subsection{The formation and evolution of planetary systems}\label{Sect:planetformation}

Planets form in disks on a time scale comparable to the cluster
formation time scale. The gaseous disk first condenses into pebbles,
which collide to form a spectrum of massive objects.  The typical
distribution of this mass spectrum is a power law with a slope of
about -1.6, which is consistent with the theoretically expected
distribution of fragments through collisions
\citep{2022SoSyR..56..338B}; low mass objects (planetesimals) are more
abundant than high-mass objects (planets). There is a break, however,
in the mass function due to the Oligarchic growth, causing a few
objects to accrete much more than the many others.

Once a planet-mass object forms, it starts opening a gap in the
gaseous and dusty disk, and migrate towards the parent star. This
migration process stops once the planet encounters a reversed density
gradient in the disk, or it runs into another planet with which it
synchronizes.  This latter process has probably happened in the
Trappist 1 system, where 7 planets are locked in 2:1 resonant
synchronous orbits \citep{2017Natur.542..456G}. The ratio here indicates
that every inner planets orbits the star twice for every outer planet
orbit. Similar resonance are observed in other planetary systems, and
even in the Solar system we observe a near-resonance in Jupiter
(orbital period of 12 years) and Saturn (orbital period of 29.4
years), in this case a near 5:2 mean motion resonance.

After the main ingredients of a planetary system have formed, the more
massive objects start to accrete the low-mass objects or eject them
from the planetary system.  In the Solar system, the (ice) giant
planets kicked out most of the planetesimals in their vicinity, and
evidence for collisions is omnipresent in the cratering on smaller
rocky objects such as the Moon and Mercury.  The majority $\apgt
60$\,\% of planetesimals between Jupiter and Neptune have been lost
this way. Some, have survived and can be found in resonant orbits
(such as the Trojans and the Greeks in an 1:1 resonance with Jupiter),
others need more time to be ejected or accreted.

Although, there are multiple short timescale relaxation processes
operating within the disk, the global two-body relaxation time scale
that dominates the evolution of the global structure is typically
longer than the lifetime of the parent star.  As a consequence, we
can, in principle, reconstruct the history of a planetary system based
on its current orbital topology.  One of these signatures is the still
relatively flat distribution of the ecliptic plane in which most
Solar system objects orbit.

This whole picture is based on the isolated formation of a planetary
system. However, stars tend to form in clustered environments, and
planetary systems form around stars.  Planetary systems then form in
the same perturbing environment, and it is expected that this environment
influences the formation and evolution of the planetary systems.

\begin{figure}
\includegraphics[width=1\linewidth]{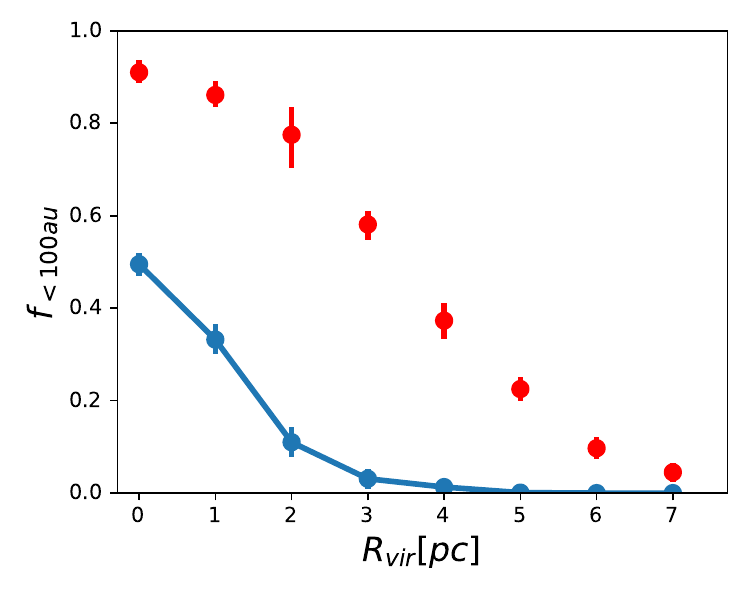}
\caption{Fraction of stars for which perturbations caused by a nearby
  passing star in the first 10\,Myr reaches to within 100\,au, as a
  function of the cluster's virial radius. We show two examples with
  100 stars each; a virialized Plummer sphere
  (blue) and a fractal distribution (with a fractal dimension of
  $1.6$, in red).
\label{fig:IntaractionProbability}
}
\end{figure}

There are several environmental processes that influence the young
planetary system.  In the earliest phase, the circum-stellar disks are
affected by the nearby stars through photo evaporation and tidal
truncation. The first process causes the disk to disperse, suppressing
the planet formation process, or even preventing planet formation
altogether.  Dynamical truncation by nearby passing stars causes disks
to be stripped off their outer rim, and introduces asymmetries such as
warps and spiral arms in the disk \citep{2017A&A...599A..91B}.
Gaseous disks are differently affected by these processes than debris
disks. These perturbed disks give rise to an additional source of
free-floating planets and planetesimals. This is an active research
field, and much is still to be explored.

\section{Theoretical Perspective}\label{Sect:Theory}

Now we have set the stage from an observational perspective in
\cref{chap1:Observations}, and roughly sketched the formation history
of the birth environments in \cref{Sect:Environment}, we can 
discuss the theoretical arguments for the origin of free-floating
objects in the Galaxy. We do that in three stages, first we discuss
free-floating stars, followed by planets and finally discuss the
interstellar asteroids.

\subsection{Planet-mass objects}\label{chap1:ffplanets}

The majority of Jupiter-mass free-floaters tends to arise from fully
packed planetary systems \citep{2023arXiv231015603C}, or as ejectees
from their birth planetary system following encounters with other
stars in the cluster \citep{2019A&A...624A.120V}.  Single Jupiter-mass
free-floating objects then originally formed in a disk around a star
to become single later in time.
The number of super-Jupiter mass free-floating objects formed in this
way are predicted to be on the order of one ($\sim 0.71$) per star
\citep{2019A&A...624A.120V}, but lower-mass free-floaters orphaned
this way, may be much more abundant;
The origin of relatively massive free-floaters through dynamical
phenomena is further complicated by the tendency for lower-mass
planets to be more prone to ejections.


On the other hand, star formation via the collapse of a molecular
cloud through gravitational instability leads to objects considerably
more massive than Jupiter.
In disks, planets tend to form with lower masses, and certainly the
more massive Jupiter-mass objects are hardest to eject.  Some of the
most massive free-floating objects then may have formed in situ, much
in the same way as stars.  Theory developed by
\cite{2006A&A...458..817W} introduces the possibility for in situ
formation of objects with Jupiter masses. In that case one probably
should consider them as failed stars rather than planets in the
classical sense.

Planets of lower mass may very well originate from dynamical
encounters between stars and young planetary systems. In particular if
the parent stars are the members of a dense cluster.  In
fig.\,\ref{fig:IntaractionProbability}, we present the fraction of
stars that experience an encounter sufficiently close to unbind
planets down to a distance of $\apgt 100$\,au from the parent star
\citep[see also][]{2018ApJ...863...45P}. The encounter probability is
calculated for a range of stellar clusters with $100$ stars each and
for various virial radii (horizontal axis). The fractions are
presented for a virialized Plummer distribution (blue), and for a
fractal distribution (red). In high density clusters the fraction of
stars that experience a close --planet-dislodging-- encounter in the
first 10\,Myr approaches unity.

\begin{figure}
\centering
\includegraphics[width=1.0\columnwidth]{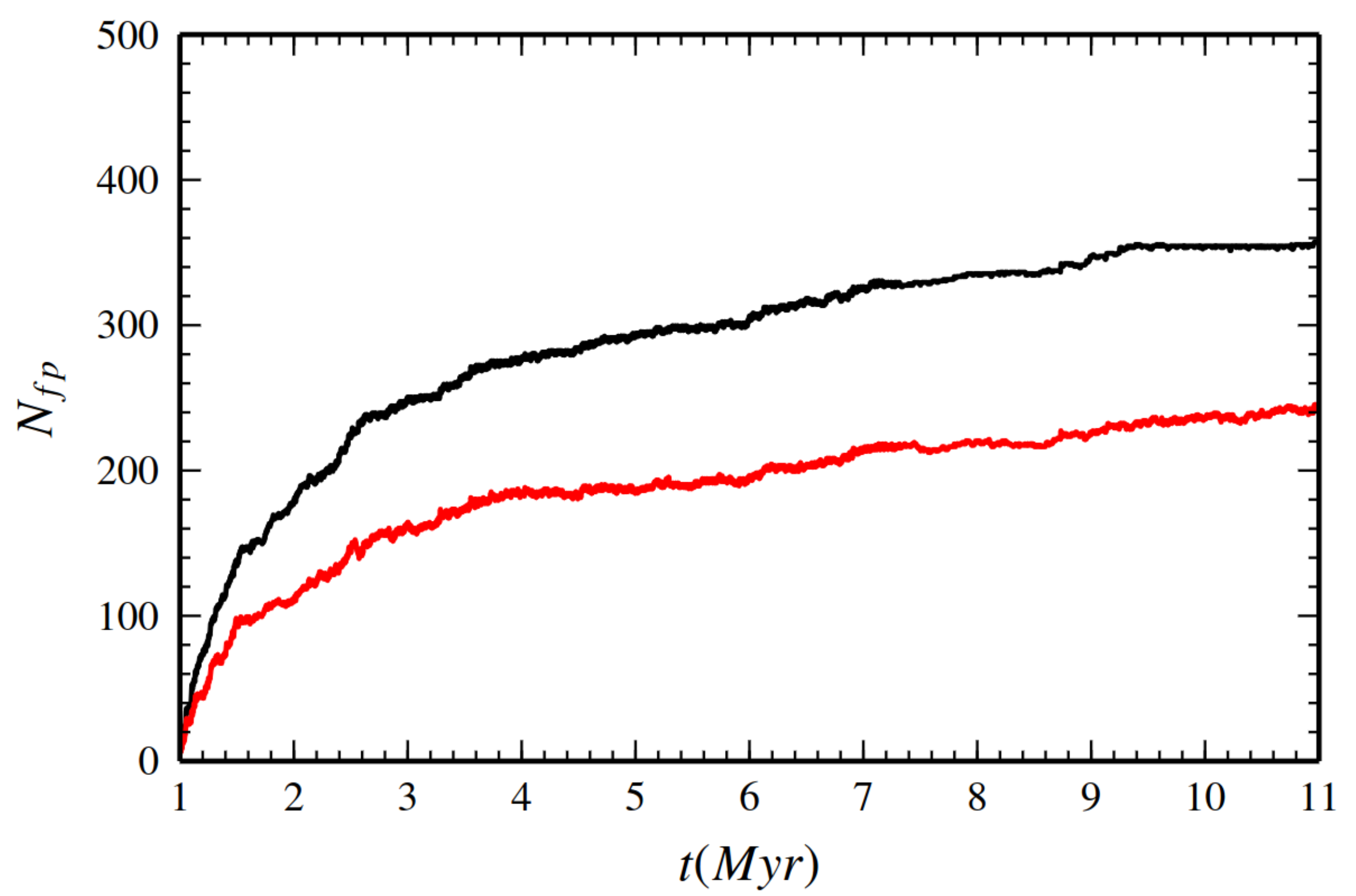}
\caption{Number of free-floating planets ($N_{fp}$) as a function of
  time. The calculations were performed using the Nemesis package in
  the AMUSE framework \citep{2018araa.book.....P} using a cluster of
  1500\,stars and 500 planetary systems in a 0.5\,pc virialized
  Plummer distribution. The solid curve (black) indicates all
  free-floating planets; the red curve indicates the subset of free
  floaters that also escape the cluster.  Figure reproduced from fig.9
  of \citep{2019A&A...624A.120V}.}
\label{fig:ffps_from_cluster}
\end{figure}

Once ejected, the free-floating planet travels first within a cluster
to eventually escape.  In \cref{fig:ffps_from_cluster} we present the
number of free floaters produced in a simulation of 1500 stars from a
Salpeter mass function and initially distributed in a 0.5\,pc
virialized fractal cluster (fractal dimension 1.6).  A total of 500
Sun-like stars received a systems of 5 planets (on average) in
circular orbits with a separation up to 400\,au in a random oriented
plane.

In the figure (\cref{fig:ffps_from_cluster}) we show the number of
planets dissociated from their parent star (in black), and those
escaping the cluster (red). In these calculations, the fraction of
free-floaters in the cluster potential is about $2/3^{\rm rd}$ of the
total number of free-floating planets. $1/3^{\rm rd}$ also escapes the
cluster to become a free-floating planet in the Galactic potential. Of
course, by the time the cluster dissolves, the bound free-floating
planets also become part of the Galactic background. We then expect
clusters to be quite rich in free-floating planets.

In simulations \cite{2019A&A...624A.120V} finds about $100$
free-floating planets among a total of 1500 stars at an age of about
1\,Myr (roughly the age of the Trapezium cluster).  The Trapezium
cluster could then, with similar initial conditions, produce $\sim
160$ free-floating planets from ionizing planetary systems.  Only
$\sim 30$\,\% of these have a mass $\apgt 1$\,\MJup\, \citep[see fig
  11 of][]{2019A&A...624A.120V}.  The eventual number of free-floating
Jupiter-mass planets from dissociated planetary systems we expect in
the Trapezium cluster then $\aplt 55$.  The majority ($\sim 90$\,\%)
of free-floating planets in the Trapezium cluster must them have
another origin.

For the large number of planet-mass objects in Upper Scorpius similar
arguments might hold, but their observations are regretfully not yet
available for further analysis.  In addition to the single
Jupiter-mass objects \cite{2023arXiv231001231P} found more than 40
paired objects, called JuMBOs. We now discuss the origin of these
objects.

\subsubsection{Jupiter-Mass Binary Objects}

Explaining the observed abundance and masses of Jupiter-Mass Binary
Objects (\jumbos) in the Trapezium cluster \citep{2023arXiv231001231P}
is not trivial.  So far, binary free-floating planetary-mass objects
have been rare, and were only discovered in tight (few au) orbits
\citep{2021ApJS..253....7K} (see also \cref{chap1:Obs_ffplanets}).
Such tight pairs could have formed as binary planets (or planet-moon
pairs) orbiting stars, before being dislodged from their parents
\citep{2016ApJ...819..125C}.  If only a few of these objects were
discovered in tight orbits, such an exotic scenario could explain
their observed abundance, but the relatively wide orbits of the
observed \jumbos\, remain hard to explain.

\begin{figure}
\begin{center}
FFC  \includegraphics[width=0.45\columnwidth]{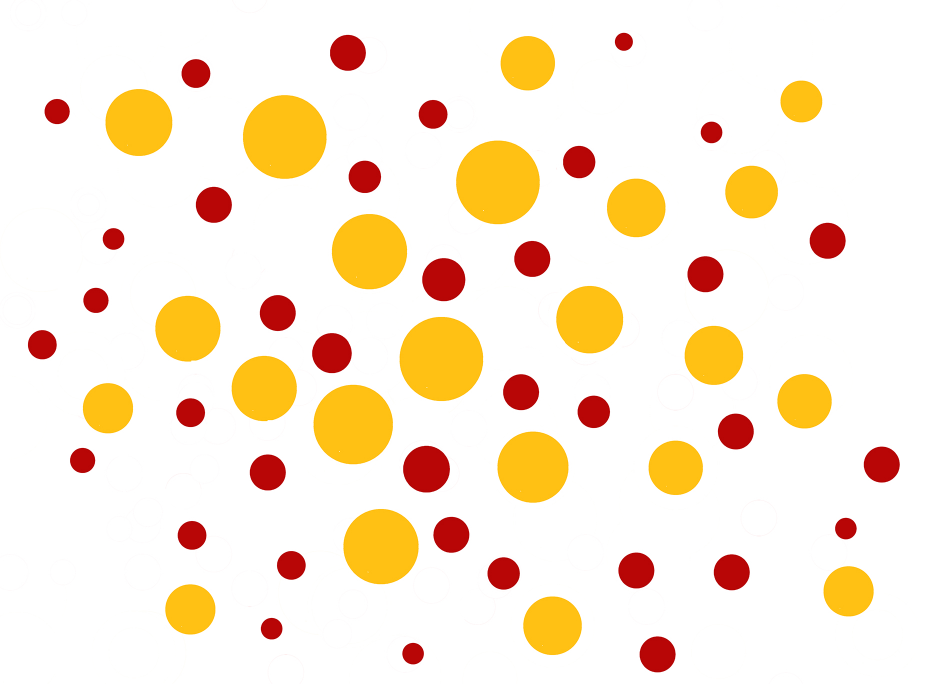}
SPP:    ~\includegraphics[width=0.40\columnwidth,angle=90]{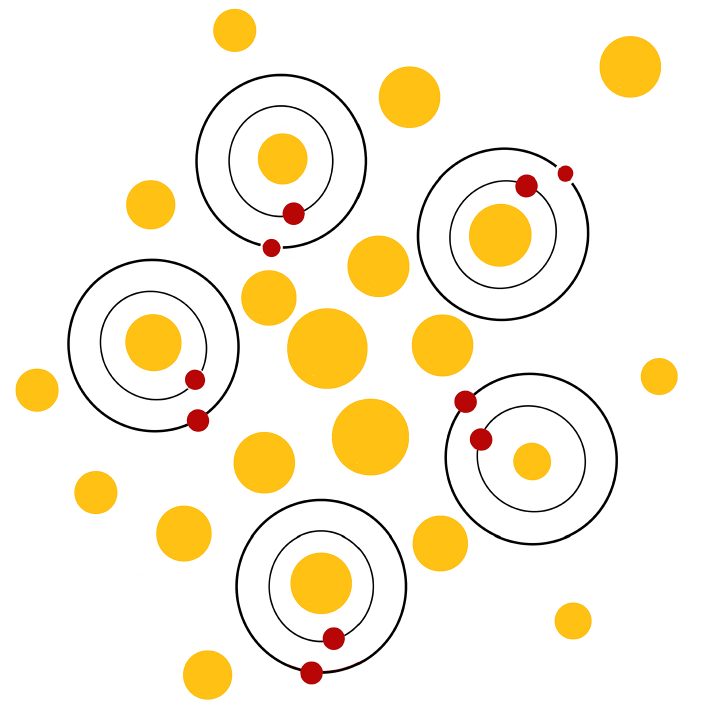}\\
SPM:    \includegraphics[width=0.40\columnwidth]{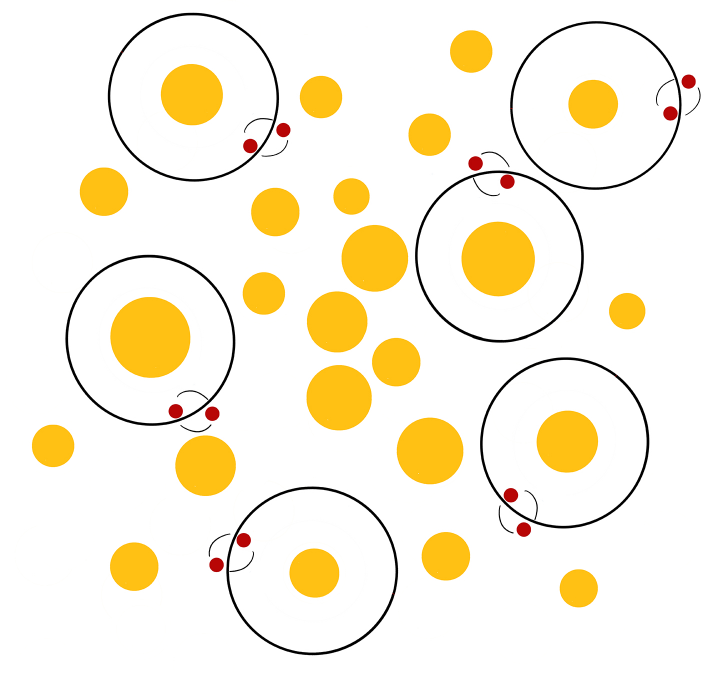}
ISF:    ~\includegraphics[width=0.45\columnwidth]{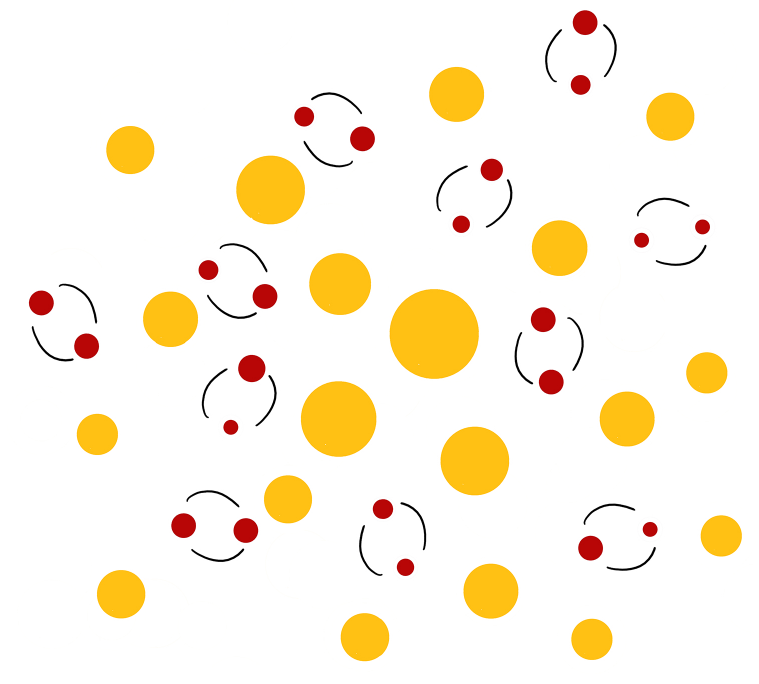}
\caption{Illustration of the four possible origins for Jupiter-mass
  binary objects (JuMBOs) in the stellar cluster.  Stars are
  represented with yellow bullets, and Jupiter-mass objects in red.
  From top left to bottom right we have (as indicated): model ${\cal
    FFC}$ for the free-floating single planets; ${\cal SPP}$ for outer
  orbiting planets; ${\cal SPM}$ as bound planet-moon pair orbiting a
  star, and model ${\cal ISF}$ in situ formation of jumbos.  This
  image was reproduced from fig.\,1 of \citep{2024ScPA....3....1P} }
\label{Fig:models}
\end{center}
\end{figure}

We naively consider four models, illustrated in \cref{Fig:models}, for
explaining the origin of these paired objects
\citep[see][]{2024ScPA....3....1P} for details):
\ind{FFC:}{Dynamically captured pair formation through interactions
  \citep[see][for a similar scenario for the formation of
    binaries]{2011MNRAS.415.1179M}.}
\ind{SPP:}{Born as a pair of planets in which stars have two outer
  orbiting Jupiter-mass planets both of which escape to form a
  free-floating pair \citep[see][for an estimate
  of the expected rate based on this scenario]{2023arXiv231006016W}.}
\ind{SPM:}{A Jupiter-mass planet-moon pair orbiting a star. The pair may
  be dislodged from its parent star by a dynamical encounter, while
  remaining bound to each other.}
\ind{ISF:}{Formed in situ as single Jupiter-mass objects and binaries.}
The first two scenarios dramatically fail to produce the number of
\jumbos\, observed in the Trapezium cluster. The third scenario, in
which a planet-moon pair is dissociated from the parent star can
produce sufficient free-floaters and \jumbos\/, but requires
unrealistically wide orbits for the planet-moon system around the
star. The last option, in situ formation, seems most promising.

An interesting alternative, proposed by Lucio Mayer (private
communication), is the formation of a \jumbo\, from a pair of gas
blobs dissociated from a circumstellar disk in a strong encounter
between two young stars.  At least one of these stars should have a
massive disk.  The characteristics observed in the Trapezium cluster's
\jumbos\, might, after the gas blobs collapse into planet-mass bodies,
resemble the soft binaries observed. However, also in this case, the
required rate is probably too high to make this scenario work.

The observed \jumbos\ and free-floating Jupiter-mass objects in the
Trapezium cluster are best reproduced if they formed in pairs and as
free-floaters together with the other stars in a smooth (Plummer)
density profile with a virial radius of $\sim 0.5$\,pc.  For fractal
(with fractal dimension 1.6) stellar density distribution this also
works, but requires relatively recent formations ($\apgt 0.2$\,Myr
after the other stars formed) or a high ($\apgt 50$\%) initial binary
fraction among free-floating Jupiter-mass objects.  This would make
the primordial binary fraction of \jumbos\, even higher than the
already large observation fraction of $\sim 8$\,\% (42/540). The
fraction of \jumbos\, will drop with time, and the lack of \jumbos\ in
Upper Scorpius \citep{2022NatAs...6...89M} could then result in its
older age, causing more \jumbos\, to be ionized. One would then also
expect that the interstellar density of Jupiter-mass objects (mostly
singles with some $\sim 2$\% lucky surviving binaries) is $\sim
0.05$\, per pc$^{-3}$ (or $\sim 0.24$ per star).  A good test for the
formation of free-floating planets and \jumbos\, would be to establish
the predicted diminishing fraction of \jumbos\, with time. Also
finding \jumbos\, that are not members of a stellar cluster would be
of considerable interest.

\subsection{Interstellar asteroids}\label{chap1:ffplanetesimals}

So far we know, planetesimals form in the circumstellar disk as a
natural by-product of the planet-formation process.  In that case,
planetesimals form around stars.  While orbiting their star (or a
planet) there are several ways in which it can become unbound. The
consequence of becoming unbound is that it becomes a free-floating
object, much like the observed objects 'Oumuamua and Borisov
(mentioned in \cref{chap1:S.SolusLapis}).  Both these objects then
probably formed around stars.

While orbiting a star, planets and other massive objects can
dynamically influence smaller objects. For example, consider the
orbits of the Voyager spacecraft --- not the fictional ones from Star
Trek, but the two real, nuclear-powered probes launched in the
1970s. These spacecraft are currently on unbound trajectories with
respect to the Sun.  The kinematics involved in launching the Voyager
probes utilized gravitational assists from the two gas giant planets
in the Solar System to enable them to escape the Sun's gravity. This
clever technique is quite common in interactions between bodies of
differing masses. In stellar dynamics, this process is known as the
equipartition of kinetic energy.
During an encounter between two bodies of different masses, they
exchange energy so that after the encounter, both objects have similar
kinetic energies. As a result, the lighter body is accelerated, while
the more massive body slows down. If the mass ratio between the two
bodies is large, the lighter body can be ejected on an escape
trajectory.

Equipartition of kinetic energy plays a crucial role in the early
asteroidal evolution of planetary systems. In this context, the mass
of the planets exceeds the mass range of the planetesimals; the latter
can essentially be treated as test particles.

\begin{figure}
\centering
\includegraphics[width=1.0\columnwidth]{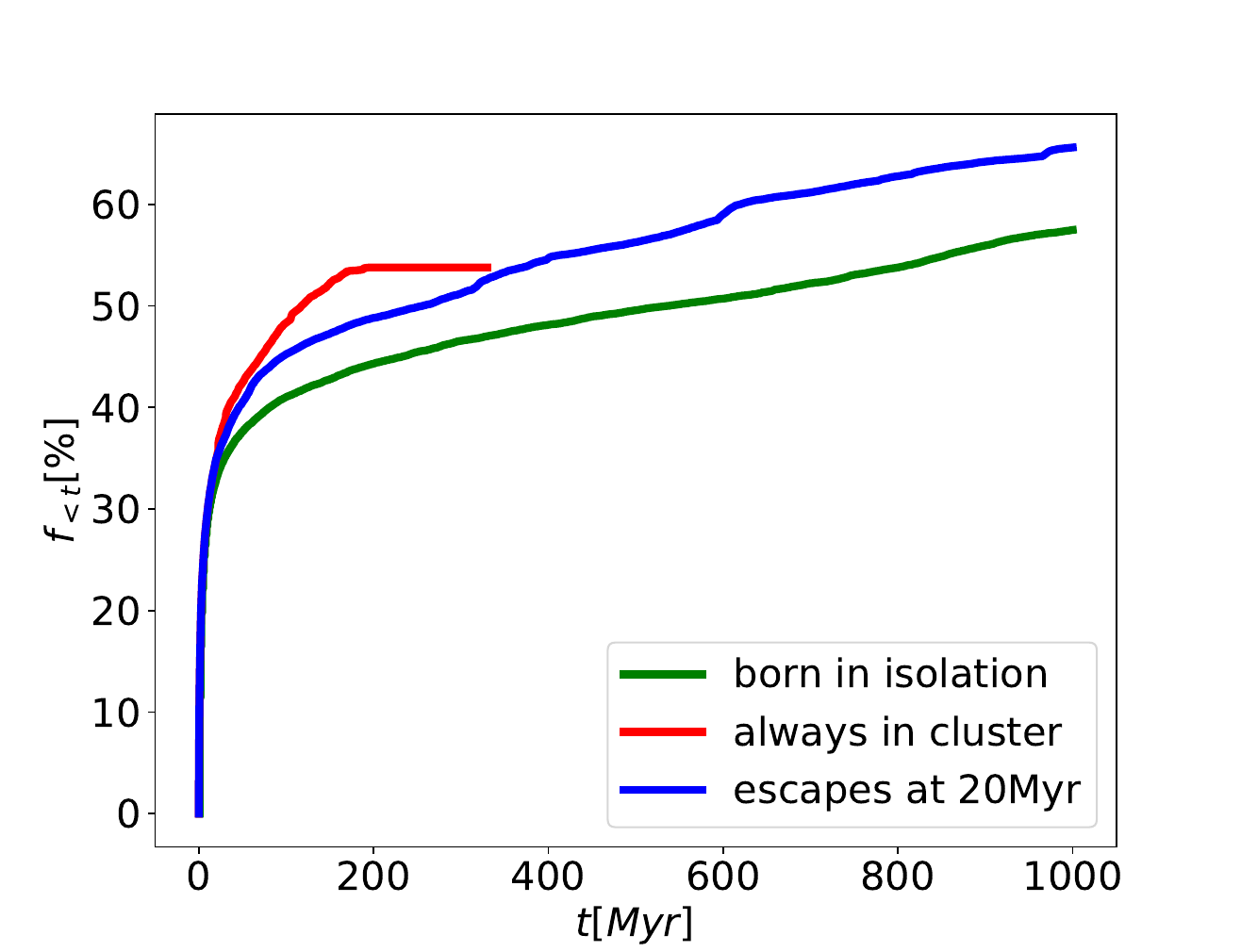}
\caption{Fraction of planetesimals that escape a young Solar system
  (star with 4 giant planets in today's orbits) for three different
  environments. The green curve gives the fraction of escapers in time
  for an isolated Solar system. the red curve (top) for the Solar
  system that spend its entire lifetime in the cluster (this curve is
  stopped at around 350\,Myr when one of the planets was lost). The
  blue curve is for the Solar system that left the cluster quietly
  after 20\,Myr.  }
\label{fig:asteroid_escape_fraction}
\end{figure}

The "{\em planetesimal bullying}" phase, lasting from 10 to 100
million years in a planetary system's evolution, results in the loss
of most primordial planetesimals. While the majority are ejected from
their stellar host, a small fraction remains in wide orbits, typically
beyond $\apgt 10^4$\,au. In the Solar System, about 50\% of
planetesimals between 1 au and 50 au were lost, with only 2 to 4\%
remaining to form the Oort cloud.
This is illustrated in \cref{fig:asteroid_escape_fraction}, where we
show the fraction of minor bodies being ejected from a
Solar-system-like planetary system (with the four (ice) giant planet
in today's orbits).  The initial system had a debris disk, composed of
test particles, between 1\,au and 50\,au. The major planets eject the
majority of these minor bodies within about 100\,Myr.  For
completeness, three model calculations are presented: 1) for the Solar
system in isolation (green), 2) the Sun born in a cluster and stays in
the cluster until one planet is lost (red), and 3) the Sun escapes the
cluster at an age of 20\,Myr (blue).

While the star orbits the Galactic center, other stars may pass,
heating of the Oort cloud, causing it to evaporate. The half
life of the Sun's Oort cloud through this process is about 4\,Gyr to
13\,Gyr \citep{2018MNRAS.473.5432H}. The amount of material released
in this way is not much compared to the earlier planetary ejection
phase, but not negligible.

When a planetary system forms within a young star cluster, close
dynamical encounters with other stars can hinder the formation of an
Oort cloud and cause both planetesimals and planets to become
unbound. Such dynamical interactions can disrupt (part of) the
circumstellar disk. In some cases, the interaction between the disk of
one star and that of another can even result in the exchange of
asteroidal or planetary material. For instance, it has been
hypothesized that the asteroid Sedna may have entered the Solar System
through this mechanism \citep{2015MNRAS.453.3157J}.

Planetesimals lost due to stellar encounters predominantly originate
from the outer regions of the disk
\citep{2021ApJ...921...90P}. Additionally, planetesimals located within
the domain of the giant planets are also susceptible to ejection,
often through gravitational slingshots by these massive planets. Only
those planetesimals that are captured in mean-motion resonances with
the giant planets or are trapped in their Lagrangian points as Trojans
(in the leading Lagrangian point $L_5$) and Greeks ($L_4$ trailing)
are likely to remain within the planetary system.

The remainder planetesimals are eventually lost once the star starts
shedding its envelope through a stellar wind, supernova or post AGB
phase \citep{2011MNRAS.417.2104V}. Some planetesimals may survive this
process \citep[][depending on their orbital parameters; semi-major
  axis, eccentricity and orbital phase]{2014MNRAS.437.1127V}.  The
remaining white dwarf (or neutron star) may keep part of the
planetesimal disk, and maybe even a some Oort cloud objects.  The
outer parts of a possible Oort cloud, however, is probably partially
lost through the orbital expansion through the stellar mass loss and
because the reduced mass of the white dwarf compared to its progenitor
causes the star's Hill radius in the Galactic potential to
shrink. Such shrinkage is proportional to the stellar mass, and a
1.6\,\MSun\, star turning into a 0.8\,\MSun\, white dwarf naturally
reduces the stars' Hill radius by about a factor of two.  These
processes cause the outer parts of the Oort cloud to become unbound
from the host star.  On the other hand, we have very little direct
knowledge of the Sun's Oort cloud, and exo-Oort clouds are purely
hypothetical.  Some white dwarfs are known to be metal rich
\citep{2024MNRAS.531L..27M}, probably due to the collisional debris
disk that rains onto its surface.  A supernova evidently causes all
planetesimals to be instantaneously lost from the star.

\begin{figure}
\centering
\includegraphics[width=1.0\columnwidth]{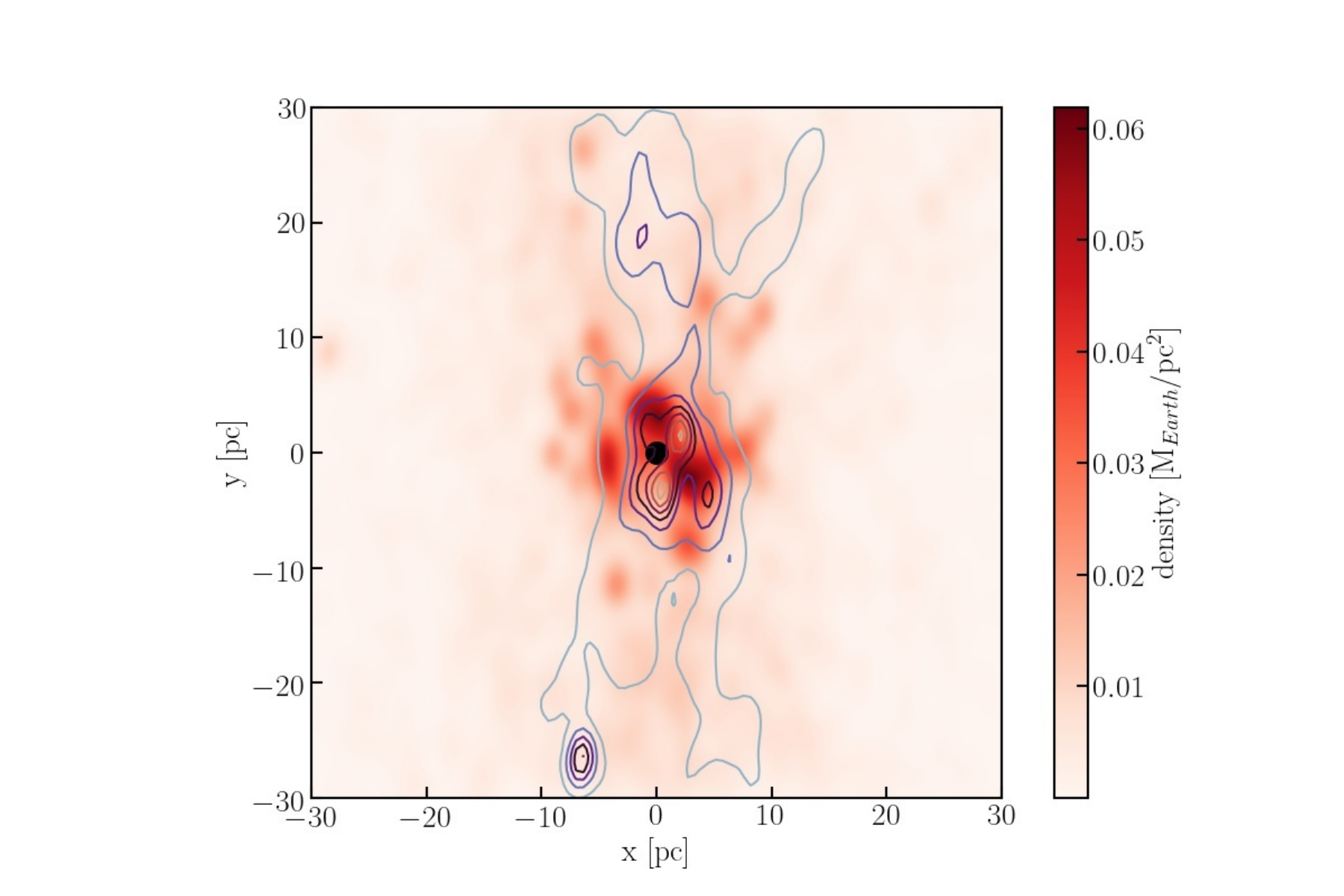}
\caption{Local projected density of asteroids within 30\,pc of the Sun
  in the Galactic $X$-$Y$ plane.  The kernel-density smoothing used
  was $\sim 0.1$\,pc.  The red shades give the projected density of
  asteroids around the Sun. The contours give the local density only
  for the asteroids that originally orbited one of the ten stars
  \citep{2021A&A...647A.136P}.  }
\label{fig:local_density_xy}
\end{figure}

The budget of free-floating objects produced by a star in this way is
almost the entire planetesimal disk, or some $10^{15}$ objects in the
range of asteroid masses. If every star in the Galaxy loses this
amount of planetesimals, and a few planets, one can envision the
entire Galaxy to be population with some $10^{26}$ Solus Lapides, and
roughly $10^{11}$ free-floating planets.

Ejected planetesimals continue to orbit the host star within the
Galactic potential, forming planetesimal streams. As the Sun moves
through the Galaxy, it may occasionally pass through one of these
streams. However, there may be no direct correlation between the Sun’s
passage through such a stream and the number of interstellar
planetesimals captured \citep{2022MNRAS.512.4062D}. In
\cref{fig:local_density_xy}, we show the projected spatial
distribution of interstellar asteroids in the Solar neighborhood,
based on data from nearby stars. This distribution assumes that all
nearby stars have planetary systems with debris disks, and that the
debris was ejected beyond the stars' Hill radii into the Galactic
potential.

\section{Outlook}

To some degree, stars could be considered free floating in the Galaxy,
although they are rarely considered as such.  A notable example of
free-floating stars are high-velocity runaway stars. These stars,
ejected during strong dynamical interactions in their parent clusters
or from supernova explosions, are unlikely to have planets or minor
bodies orbiting them. Similarly, hyper-velocity stars, accelerated by
supermassive black holes,
are not expected to be orbited by planets or planetesimals.

Non-stellar free-floating objects are probably common but challenging
to detect.  Still, several free-floating planets have been discovered
using various methods. The debate on their frequency and
characteristics is lively, with estimates suggesting that
free-floating planets in the Galaxy might outnumber stars.  Recently,
large populations of free-floating planets were discovered in the
direction of the Trapezium cluster and $\sigma$\, Orionis.  It is not
clear, at this point, if those formed as planets (orbiting a star) or
in situ (like a star).  An unexpectedly high fraction of free-floating
planets found in the direction of the Orion Trapezium cluster appear
in unusually wide pairs, which challenges traditional formation models
but aligns with the idea of in-situ formation. These objects might be
classified similarly to stars, as they may have formed in a comparable
manner despite being too small for internal fusion.

The intriguing population of interstellar planetesimals, or Solus
Lapides, is represented by only two known examples so far. Despite
their differences, theorists estimate their total number in the Galaxy
could exceed the number of stars by factors of $10^{11}$ or
$10^{12}$. These planetesimals may impact Galactic processes,
including star and planet formation. It is possible that several such
objects are currently present in the Solar System without our
knowledge. The population of free-floating planets is probably
considerably smaller, with only ${\cal O}(1)$ per star.  Free-floating
objects are an important component of the Galactic
population. Studying their dynamics and whereabouts can provide
valuable insights into various astronomical processes, such as planet
formation and cluster dynamics. The primary challenge remains their
detection.

\begin{ack}[Acknowledgments]

  It is a pleasure to thank the editor Dimitri Veras for suggesting
  this topic, and suggestions on the manuscript.  I also like to thank
  Shuo Huand and Erwan Hochart for discussions.
  
\end{ack}



\input{./disks_and_planets.bbl}

\end{document}